\documentclass[onecolumn,showkeys,preprintnumbers,aps,a4paper,amssymb,prd,superscriptaddress,nofootinbib]{revtex4-2}
\usepackage{comment}
\usepackage{graphicx}
\usepackage{epsf}
\usepackage{bm}
\usepackage{amsmath}
\usepackage{amsfonts}
\usepackage{amssymb}
\usepackage{epstopdf}
\usepackage{color}
\usepackage[dvipsnames]{xcolor}
\usepackage{verbatim}
\usepackage{multirow}
\usepackage{soul}
\usepackage{physics}
\usepackage{bm}

\usepackage[width=0.00cm, height=0.00cm, left=1.50cm, right=1.50cm, top=2.00cm, bottom=2.00cm]{geometry}
\usepackage{microtype}
\usepackage{lmodern}

\usepackage[colorlinks = true,
linkcolor = teal,
urlcolor  = teal,
citecolor = teal,
anchorcolor = blue]{hyperref}
\usepackage[capitalize]{cleveref}
\usepackage[normalem]{ulem}
\usepackage{enumitem}
\usepackage{booktabs}

\usepackage{lipsum}

\makeatletter\let\expandableinput\@@input\makeatother

\begin{document}
\begin{center}
		\vspace{0.4cm} {\large{\bf Alleviating the Hubble Tension with Logarithmic Dark Energy: Constraints on the $w_{log}$CDM Model
}} \\
		\vspace{0.4cm}
		\normalsize{Saurabh Verma$^1$, Archana Dixit$^2$, Anirudh Pradhan$^3$, M. S. Barak$^4$ }\\
		\vspace{5mm}
		
        \normalsize{$^{1,4 }$ Department of Mathematics, Indira Gandhi University, Meerpur, Haryana 122502, India }\\ 
        \normalsize{$^{2}$ Department of Mathematics, Gurugram University, Gurugram, Harayana, India\\

		\normalsize{$^{3 }$ Centre for Cosmology, Astrophysics and Space Science (CCASS), GLA University, Mathura-281406, Uttar Pradesh, India}\\ 

		\vspace{2mm}
		$^1$Email address: saurabh.math.rs@igu.ac.in\\
            $^2$Email address: archana.ibs.maths@gmail.com\\
		$^3$Email address: pradhan.anirudh@gmail.com\\
        $^4$Email address: ms$_{-}$barak@igu.ac.in\\}
\end{center}

 
\maketitle
%
{\bf Abstract}:
 Observational constraints are considered on a $w_{log}$CDM model of the 
dark energy equation of state, $w_{d}(z) = w_{0} + w_{a}\left( \frac{\ln(2+z)}{1+z} - \ln 2 \right)$, using the most recent cosmological datasets including DESI Baryon Acoustic Oscillation (BAO) measurements, Big Bang Nucleosynthesis (BBN) priors, Cosmic 
Chronometer (CC) observations, and Pantheon Plus (PPS) Type Ia supernovae. From the combined DESI BAO+BBN+CC+PPS dataset, we obtain
$H_0 = 71.02 \pm 0.66~\text{kms}^{-1}\text{Mpc}^{-1}$, $\Omega_m = 0.2863 \pm 0.0080,$ $w_0 = -0.875 \pm 0.066,$ 
$w_a = -0.69^{+0.37}_{-0.32},$ at the 68\% and 95\% confidence levels, indicating a preference for phantom dark energy with mild evidence for temporal evolution. The Hubble constant obtained from our model is closer to the local SH0ES measurement than the standard $\Lambda$CDM prediction, partially easing the Hubble tension. We perform extensive parameter-space exploration revealing correlations between $w_0$, $w_a$, and $H_0$, showing that dynamical dark energy models can fit 
higher values of the Hubble constant. The reconstructed deceleration parameter 
$q(z)$ shows the transition from deceleration to acceleration at $z \sim 0.6$--$0.7$, while the equation-of-state reconstruction remains 
consistent with a cosmological constant across the observed redshift range. A model comparison using information criteria indicates that the $w_{log}$CDM model remains statistically competitive with $\Lambda$CDM.

\section{Introduction}

In the final years of the 1990s, thorough studies of distant Type Ia supernovae (SNe Ia) at high redshift yielded the groundbreaking discovery that our Universe is expanding at an accelerating rate \cite{ref1,ref2}. This result has since been continually confirmed by new developments in observational cosmology over subsequent years. Various theoretical models have been proposed to explain this accelerating expansion, most of which can be categorized into two main classes. The first class interprets acceleration by invoking a mysterious component with negative pressure, widely known as dark energy (DE). The second class of theories attempts to explain observations by incorporating modifications to Einstein's general relativity at cosmological scales \cite{ref3,ref4,ref5}. Working within general relativity, the most natural candidate for dark energy is the cosmological constant $(\Lambda )$, representing approximately $70\%$ of the total energy content \cite{ref6,ref7,ref8}. When this cosmological constant is combined with cold dark matter, the framework  successfully describes the current epoch of the cosmic acceleration. Despite its empirically satisfactory results, the $\Lambda$CDM paradigm poses serious theoretical problems \cite{ref9,ref10,ref11,ref11a}.\\

Beyond such theoretical concerns, there are several observational discrepancies in the $\Lambda$CDM model's determination of the basic cosmological parameters. For example, Baryon Acoustic Oscillation measurements from the Lyman-$\alpha$ forest \cite{ref12} tend to prefer a lower matter density parameter compared to values extracted from Cosmic Microwave Background analyses. Disagreement can also be found among the different observations of large-scale structure \cite{ref13}. Another, perhaps even more serious problem, is that the values of the Hubble constant $(H_0)$ from local distance ladder measurements are inconsistent with the values determined from Planck CMB data \cite{ref14}. More precisely, the value of $H_0= 67.4 \pm 0.5$ km/s/Mpc \cite{ref15} from Planck CMB observation within $\Lambda$CDM is inconsistent with the value  $H_0=74.03 \pm 1.42$ km/s/Mpc \cite{ref16} obtained from Cepheid-calibrated Type Ia supernovae. Although $\Lambda$CDM provides an elegant explanation for the acceleration of the universe, it nevertheless faces some fundamental theoretical puzzles, such as fine-tuning and coincidence problems. These reasons have motivated the extensive investigation of alternative scenarios, either based on modified gravity theories or dynamic dark energy. Finally, a number of model-independent cosmographic analyses, independent of any parameterization, present in recent literature show that, overall, the significant discrepancies between the predictions of $\Lambda$CDM and the model-independent results provide strong motivations for testing possible deviations from the standard paradigm \cite{ref17,ref18,ref19,ref20,ref21}.\\

Several alternative models to $\Lambda$CDM have been developed to describe the dark energy. Although the cosmological constant is still the most natural candidate \cite{New1}, other proposals include quintessence fields \cite{New2}, phantom energy \cite{New3}, and various time-dependent parameterizations. Among the latter, the most commonly used are the Chevallier–Polarski–Linder (CPL) formulation \cite{New4,New5}and the Barboza–Alcaniz (BA) approach \cite{New6}. Each of these was designed to capture possible deviations from a constant equation of state at different cosmological epochs. In addition to the $\Lambda$CDM model, several dark energy models have already been considered, such as the $w$CDM model, where dark energy is considered as a perfect fluid with a constant equation of state parameter $w$ different from $-1$, together with many dynamical scenarios presented in the literature \cite{ref22,ref23,ref24,ref25,ref26,ref27,ref28,ref28a,ref28b,ref28c}. A productive approach toward the investigation of dynamical dark energy involves adopting  a redshift-dependent parameterization for $w(z)$. Given the ignorance about the nature of dark energy, several functional forms for $w(z)$ have already appeared in the literature \cite{ref31,ref32,ref33,ref34,ref35,ref36,ref37,ref37a}.\\

Because scalar field models are characterized by various forms of potential and testing each one poses practical problems, dark energy evolution can be parameterized in a way that represents large classes of models. The most common form of parameterization is to parameterize the equation of the dark energy state, $(w=\frac{p}{\rho})$. To determine the dynamical evolution of these models, we use the EoS parameter $w$ as a function of redshift from the extensive list of its parametrizations presented in the literature \cite{ref38,ref39,ref40,ref41,ref42,ref43,ref44,ref45,ref46,ref47,ref48,ref49}. We also compared the behavior of the EoS parameter with a number of other popular parameterizations: CPL \cite{ref50,ref51,ref52,ref53} and BA \cite{ref54}, tested against observational data along with the standard $\Lambda$CDM model and constant equation of state models. Constraints on the equation of state parameters have been obtained from several cosmological probes such as weak gravitational lensing power spectra, BAO, CMB anisotropies, and SNe Ia. The use of these complementary datasets is important for furthering our understanding of the DE physics. The tightest bounds on the EoS values thus far are provided by supernovae observations, DESI BAO measurements, and Cosmic Chronometers (CC) data. The conventional $\Lambda$CDM model is often defined by a two-parameter parameterization given by $(w_0, w_a)$, where $w_0$ is the value of the equation of state today and $w_a$ is the variation of the equation of state with time, which may or may not be supported by observations.\\

In this study, we investigated several well-established dark energy parameterizations. We start our analysis with the Chevallier–Polarski–Linder (CPL) parameterization, for which several weaknesses when dealing with high redshift $z$ have been discussed in the literature \cite{ref38,ref39,ref40}. To bypass these issues, several alternative parameterizations that preserve the same physical meaning of $w_0$ and $w_a$ have been proposed. We studied a parallel $w_{log}$CDM model \cite{ref61}. These frameworks model dark energy at low and high cosmological scales, exhibiting large evolution at low redshifts where the above CPL form cannot be easily extended. Our goal is to constrain the bi-dimensional dark energy $w_{log}$CDM model with the latest available observational evidence, including PantheonPlus$\&$SH0ES, DESI BAO, and CC measurements, and determine which one yields better constraints. Before proceeding, it is important to clarify that the $w_{log}$CDM model considered in this study is a phenomenological parameterization of the dark-energy equation of state. The logarithmic form of $w_d(z)$ is not derived from an underlying scalar-field Lagrangian or a modified-gravity framework. Rather, it is introduced as an effective and flexible description aimed at capturing possible smooth deviations from a constant equation of state over the observationally relevant redshift range, while remaining well behaved at both low and high redshifts. The structure of the paper is as follows: in Sec. II we briefly introduce the theoretical framework and discuss the dark energy $w_{log}$CDM model in this study. In Section III, we describe
the datasets used for our analysis and discuss the methodology adopted. Section IV presents our main results and a discussion of the $w_{log}$CDM model. A comparison of the model is presented in Section V. Finally, Section VI summarizes our findings and suggests future possibilities for this research.

\section{Evolution of background and $w_{log}$CDM model}\label{sec2}

\textbf{ Background Evolution}
The evolution of a spatially flat FRW universe with a homogeneous and isotropic background is usually governed by
\begin{equation}
H^{2} = \frac{8\pi G}{3} (\rho_{m} + \rho_{de} )
\label{eq1}
\end{equation}
Here, $H = \dot{a}/a$ denotes the Hubble parameter, and $\rho_{i}$ represents the energy density of the $i$-th component, where $i$ corresponds to the dark matter $(m)$ and dark energy $(de)$. We also assume that each fluid component is minimally coupled to gravity and that no interaction occurs between them. Consequently, all the components evolve independently, and their energy densities satisfy the following continuity equations:

\begin{align}
\dot{\rho}_{m} + 3H\rho_{m} &= 0, \label{eq2} \\
\dot{\rho}_{d} + 3H(1 + w_{d})\rho_{d} &= 0. \label{eq3}
\end{align}
where the dot gives the meaning of the derivative relative to cosmic time $t$. Using Eqs.~(\ref{eq2})--(\ref{eq3}), one can obtain 
$\rho_{m} = \rho_{m0} (1+z)^{3}$,  
and $\rho_{d} = \rho_{d0} g(z)$, 
where $g(a)$ is defined as follows:
\begin{equation}
g(a) = \exp\left[3 \int_{0}^{z} \frac{1 + w_{d}(z')}{1+z'} dz'\right].
\label{eq4}
\end{equation}
Hence, considering the definitions 
$\Omega_{i} = \frac{8\pi G \rho_{i}}{3H^{2}}$
and normalized Hubble parameter 
$E(z) = \frac{H(z)}{H_{0}}$, 
Eq.~(\ref{eq1}) can be written as follows:
\begin{equation}
E^{2}(z) = [\Omega_{m0}(1+z)^{3} + \Omega_{r0}(1+z)^{4}+ \Omega_{d0}g(z)].
\label{eq5}
\end{equation}

 To obtain $E(z)$ related to the parameterizations considered in this study, $w_{d}(z)$ of these parameterizations can be replaced by Eq.~(\ref{eq4}), and finally insert the result in Eq.~(\ref{eq5}).\\
\textbf{$w_{log}$CDM model}\\

We consider the parameterization of the logarithm of the EoS parameter $w_{d}$ of the effective fluid in terms of time $z$, that is,

\begin{equation}
w_{d}(z) = w_{0} + w_{a} \left( \frac{\ln(2 + z)}{1 + z} - \ln 2 \right),
\label{eq6}
\end{equation}
where $w_{0}$ also denotes the value of $w_{d}(z)$ at the present time, and $w_{a}$ is another parameter characterizing the evolution of $w_{d}(z)$:
It is noteworthy that in order to keep $w_{0}$ as the current value of $w_{d}(z)$, a minus $ln 2$ is retained in the bracket.

We emphasize that the logarithmic form of the dark-energy equation of state should be interpreted as a phenomenological parameterization rather than a fundamental description. The $w_{log}$CDM model is not constructed from a specific scalar-field potential, modified-gravity theory, or a microscopic dark-energy mechanism. Instead, it provides an effective framework for exploring possible time variation in the dark-energy sector in a model-independent and observationally driven manner.\\

Using equations \eqref{eq4} and \eqref{eq6} in \eqref{eq5}, we obtain $E(z)$ as:

\begin{equation}
E(z)=\frac{H(z)}{H_0} =  \sqrt{ \Omega_{m0}(1+z)^3 + (1 - \Omega_{m0})(1+z)^{3(1 + w_0 - w_a \ln 2)} 
\exp\left[ 3w_a \left( -\frac{\ln(2+z)}{1+z} + \ln\frac{4(1+z)}{2+z} \right) \right] }.
\label{eq7}
\end{equation}
where $\Omega_{m0}$ is the matter density parameter, and  we have $\Omega_{d0}=1-\Omega_{m0}$ as the DE parameter; hence, we obtain the final expression, in which the Hubble parameter $(H)$ is explicitly defined as a function of redshift $(z)$ as:

\begin{equation}
{H(z)}={H_0} \sqrt{ \Omega_{m0}(1+z)^3 + (1 - \Omega_{m0})(1+z)^{3(1 + w_0 - w_a \ln 2)} 
\exp\left[ 3w_a \left( -\frac{\ln(2+z)}{1+z} + \ln\frac{4(1+z)}{2+z} \right) \right] }.
\label{eq8}
\end{equation}

\textbf{Physical Motivation}
Despite its phenomenological nature, parametrizations such as $w_{log}$CDM can effectively mimic the behavior of a wide class of underlying dark-energy scenarios. For example, scalar-field models with slowly varying potentials, k-essence theories, or effective-field-theory descriptions of cosmic acceleration may lead to a smooth and non-trivial redshift evolution of the equation of state, which can be approximated by logarithmic or similar functional forms over the range probed by current observations. Thus, the $w_{log}$CDM model serves as a useful proxy for testing deviations from $\Lambda$CDM without committing to a specific fundamental theory.

\section{Datasets and Methodology}
   
\subsection{\textbf{Type Ia supernovae and Cepheid:}}

Type Ia supernovae play an important role in the development of the conventional model of the universe. From SNe Ia, one can obtain useful distance modulus measurements, a constraint on the late-time expansion rate, or the uncalibrated luminosity distance $H_0d_L(z)$. The theoretical apparent magnitude $m_B$ from a supernova at redshift $z$ is given by
\begin{eqnarray}
\label{distance_modulus}
m_B = 5 \log_{10} \left[ \frac{d_L(z)}{1{\rm Mpc}} \right] + 25 + M_B,
\end{eqnarray}
where $M_B$ denotes the absolute magnitude. The distance modulus ($\mu(z)$) can be expressed as $\mu(z) = m_{B} - M_{B}$. In a flat cosmology, the luminosity distance is defined as:
\begin{equation}
\label{eq:dl}
d_L(z) = (1+z)\int_0^{z}\frac{dz^{\prime}}{H(z^{\prime})}.
\end{equation}

In this work, we utilized the SNe Ia distance modulus measurements from the Pantheon Plus sample \cite{ref62}. We refer to this dataset as PP, and it consists of $1701$ light curves that represent $1550$ unique supernovae Ia in the redshift range of $z$ $\in$ $[0.01, 2.26]$. We further put constraints on $H_0$ and $M_B$ by incorporating SH0ES Cepheid host distance anchors \cite{ref62} into our analysis, henceforth referred to as the PPS.

\begin{figure}[hbt!]
    
  \includegraphics[width=0.7\linewidth]{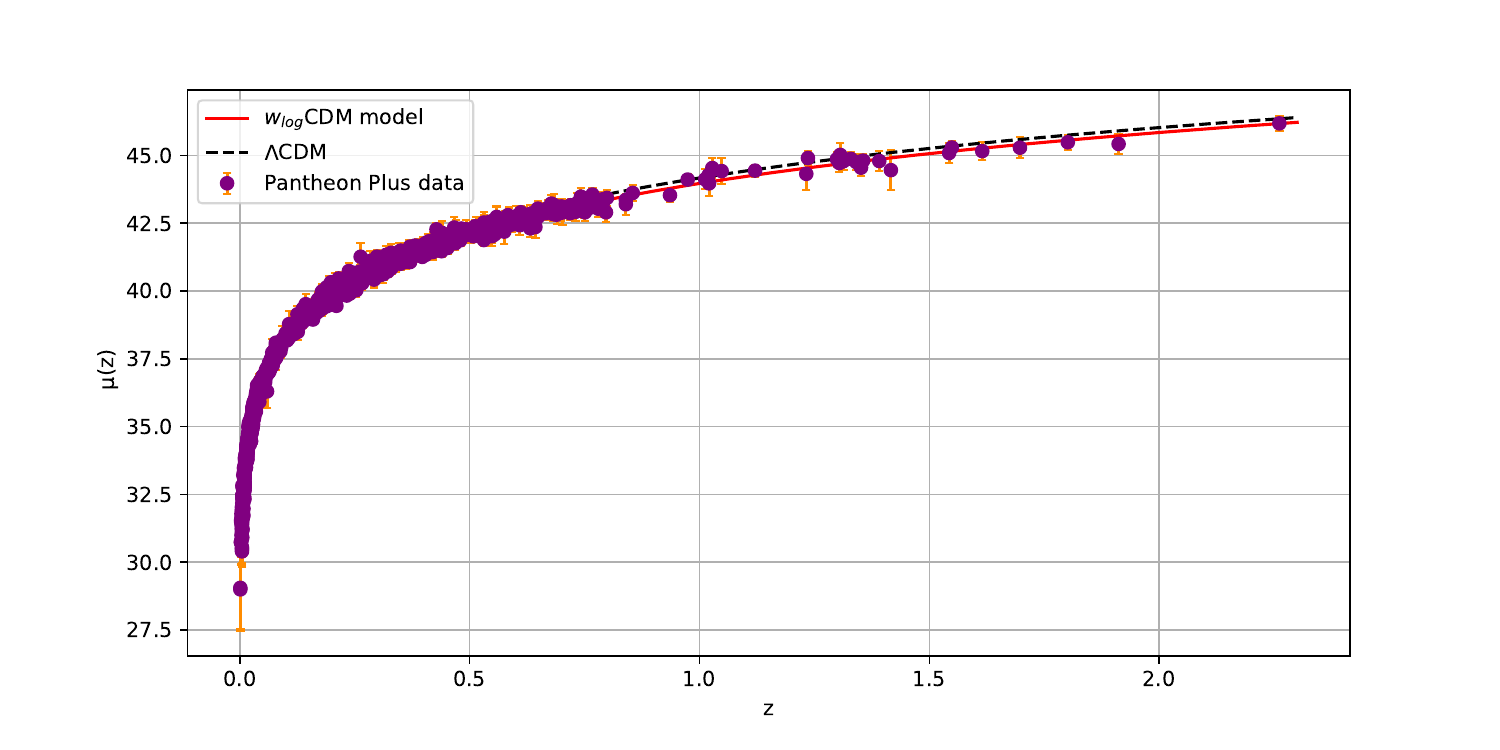}
    \caption{Distance modulus $\mu(z)$ as a function of redshift for the Pantheon Plus Type Ia supernova sample. The purple data points with error bars represent the observed $\mu(z)$ measurements. The solid red curve corresponds to the best-fit $w_{\log}$CDM model based on the logarithmic parameterization of the dark-energy equation of state, while the dashed black curve denotes the standard $\Lambda$CDM prediction.}

    \label{f1}
\end{figure}

\subsection{\textbf{Cosmic Chronometer}}

The CC method ia a powerful observational method for probing the expansion history of the universe, and was first proposed in \cite{ref63}. In this approach the Hubble parameter is estimated by analyzing the age difference of passively evolving early type galaxies. In the context of the FRW cosmological model, the Hubble parameter can be defined as $ H(z) = -\frac{1}{1+z} \frac{dz}{dt} $ implying that, if one can measure how the redshift \( z \) varies with cosmic time \( t \), that is, the derivative \( dz/dt \), one can obtain the Hubble parameter \( H(z) \) directly. This implies an emphasis on the fact that it is the variation with cosmic time \( dt \) of the variation in the redshift \( dz \) that is relevant. This provides us with the Hubble parameter \( H(z) \). Therefore, it is important to use CC data to further constrain cosmological models. A detailed elaboration of the method, its realization, possible sources of errors, and other related caveats are presented in \cite{ref64}. Throughout this study, we  use the dataset on $H(z)$ discussed in \cite{ref64, ref65}. The CC method has obtained the measurement of \( H(z) \) 33 times for the range of the redshift \( 0 < z < 2 \), which encompasses approximately 10 Gyr of cosmic history \cite{ref64, ref65, ref66, ref67, ref68}. Similarly, we also include the current local determination of the Hubble constant $( H_0)$, as obtained in \cite{ref69} with a precision of $2.4\%$: $H_{0} = 73.02 \pm 1.79 \, \text{km/s/Mpc}$.

\begin{figure*}[hbt!]
    
  \includegraphics[width=0.7\linewidth]{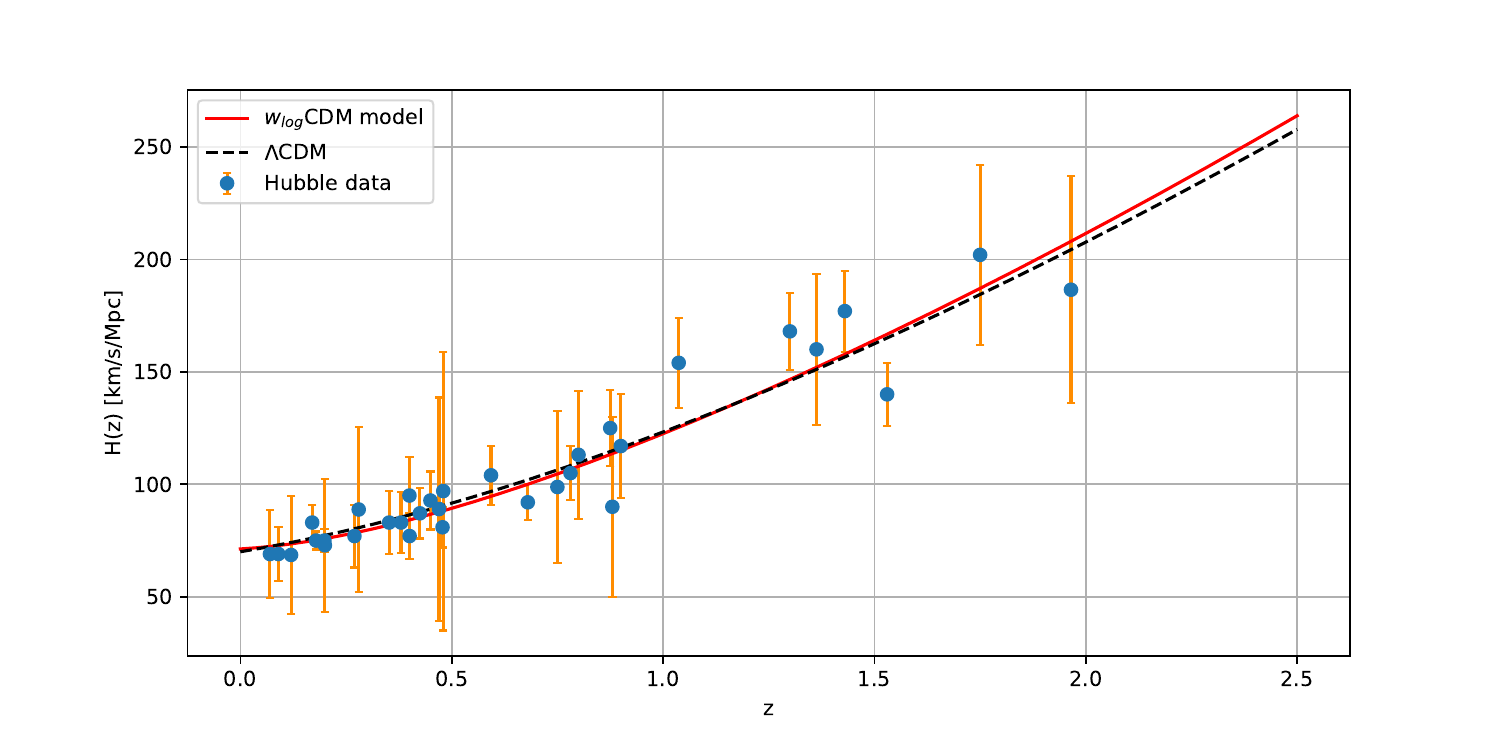}

    \caption{Best-fit reconstruction of the Hubble parameter $H(z)$ for the logarithmic dark-energy model using observational Hubble data. The orange points with error bars represent the measured $H(z)$ values from cosmic chronometer observations. The solid red curve corresponds to the best-fit $w_{\log}$CDM model, while the dashed black curve shows the standard $\Lambda$CDM prediction for comparison. }

    \label{f2}
\end{figure*}

\subsection{\textbf{DESI BAO}}
We used the BAO data released by the DESI collaboration, which includes observations from galaxies, quasars~\cite{ref70}, and Lyman-$\alpha$ tracers~\cite{ref71}, as summarized in Table I of Ref.~\cite{ref72}. These measurements cover the redshift interval $0.1 < z < 4.2$ and consist of both isotropic and anisotropic BAO observations, grouped into seven redshift bins\label{tableI}. The isotropic BAO data are expressed in terms of $D_{\mathrm{V}}(z)/r_{\mathrm{d}}$, where $D_{\mathrm{V}}$ represents the angle-averaged distance scaled by the comoving sound horizon at the drag epoch. The anisotropic BAO measurements provide $D_{\mathrm{M}}(z)/r_{\mathrm{d}}$ and $D_{\mathrm{H}}(z)/r_{\mathrm{d}}$, with $D_{\mathrm{M}}$ denoting the comoving angular diameter distance and $D_{\mathrm{H}}$ representing the Hubble horizon. We additionally accounted for the correlation between the $D_{\mathrm{M}}/r_{\mathrm{d}}$ and $D_{\mathrm{V}}/r_{\mathrm{d}}$ measurements. This data collection is referred to as  the \texttt{DESI} sample.

\subsection{\textbf{BBN}} In this section, we apply the BBN analysis to the case of $f(R)$ gravity. It is important to note that, in general, within modified gravity theories, inflation is not directly driven by the inflaton field. Instead, inflationary dynamics are embedded within gravitational modification itself. Big-bang nucleosynthesis (BBN) is the most reliable probe for the early universe, based on well-understood Standard Model physics \cite{ref76,ref77}. Abundance of the light elements $D$, $^3He$, $^4He$, and $^7Li$ predicted in the "first three minutes" are consistent with observational data, supporting the hot big-bang theory. Thus, the BBN imposes strong limits on hypothetical deviations from mainstream cosmology  and novel physics beyond the mainstream model. Big Bang Nucleosynthesis (BBN) is considered with state-of-the-art assumptions, which consist of measurements of the primordial abundances of helium, $Y_{P}$\cite{ref73}, and deuterium measurement, $y_{DP}$ = $10^{5}n_{D}/n_{H}$, obtained in \cite{ref74}. All cosmological observables were computed using the CLASS \cite{ref78,ref79}. To derive bounds for the proposed scenarios, we modified the efficient and well-known cosmological package MontePython \cite{ref80}.  

\section{Result and Discussion}
Here, we discuss the observational constraints on the $w_{log}$CDM model of the dark energy equation of state, $w_{d}(z) = w_{0} + w_{a} \left( \frac{\ln(2 + z)}{1 + z} - \ln 2 \right)$, using DESI BAO measurements in combination with the BBN, CC, PP, and PPS datasets. The $w_{log}$CDM model allows us to capture the possible evolution in the dark energy equation of state via two parameters: present-day value, $w_0$, and $w_a$, which characterizes the evolution with redshift. Table \ref{t1} summarizes the marginalized constraints on key cosmological parameters obtained from three dataset
combinations: DESI BAO+BBN+CC, DESI BAO+BBN+CC+PP, and DESI BAO+BBN+CC+PPS, comparing
the results from both the $w_{log}$CDM model and the $\Lambda$CDM model at 68\% and 95\% confidence levels. For the $w_{log}$CDM model extension allowing time-varying dark energy, the Hubble constant is constrained to $H_0 = 65.5^{+2.5}_{-2.9}$ km s$^{-1}$ Mpc$^{-1}$ for DESI BAO+BBN+CC, $H_0 = 67.23 \pm 0.84$ km s$^{-1}$ Mpc$^{-1}$ for DESI BAO+BBN+CC+PP, and
$H_0 = 71.02 \pm 0.66$ km s$^{-1}$ Mpc$^{-1}$ for DESI BAO+BBN+CC+PPS. The inclusion of PPS data significantly tightens the constraints and shifts the central value of $H_0$ upward, bringing it closer to the local measurement from SH0ES \cite{ref81} of $H_0 = 73.04 \pm 1.04$ km s$^{-1}$ Mpc$^{-1}$. However, it is important to emphasize that  Hubble tension is not fully resolved within the $w_{log}$CDM framework. A residual discrepancy at the level of approximately $\sim 2\sigma$ persists, indicating that the proposed model alleviates but does not completely solve the Hubble tension. With $\Omega_m$, all the combinations yield a consistent value in the range of $0.287$ to $0.311$, while $M_B$ has some variations with different combinations, which reflects differences mainly in the distance calibration. The dark energy equation of the state parameters revealed interesting behavior across different dataset combinations. For the present-day value, we obtain $w_0 = -0.68 \pm 0.21$ (DESI BAO+BBN+CC), $w_0 = -0.845 \pm 0.058$ (DESI BAO+BBN+CC+PP), and $w_0 = -0.875 \pm 0.066$ (DESI BAO+BBN+CC+PPS), all indicating preference for phantom dark energy though remaining consistent with the cosmological constant ($w=-1$) within uncertainties. The evolution parameter shows $w_a = -0.82^{+0.66}_{-0.54}$ (DESI BAO+BBN+CC), $w_a = -0.36^{+0.33}_{-0.26}$ (DESI BAO+BBN+CC+PP) and $w_a = -0.69^{+0.37}_{-0.32}$ (DESI BAO+BBN+CC+PPS), with all values consistent with zero evolution within $2\sigma$, although showing a mild preference for negative $w_a$.  

\begin{table*}[hbt!]

\caption{
The "$w_{log}$CDM model, and $\Lambda$CDM" models acquired from the DESI BAO+BBN+CC, DESI BAO+BBN+CC+PP and DESI BAO+BBN+CC+PPS datasets have constraints at 68$\%$ and 95$\%$ CL on a few chosen parameters.}
\vspace{0.5cm}
\label{t1}

\centering
\resizebox{\textwidth}{!}{  
\begin{tabular}{| c | c | c | c |}
\hline
Data & DESI BAO+BBN+CC & DESI BAO+BBN+CC+PP & DESI BAO+BBN+CC+PPS \\
\hline
Model & $w_{log}$CDM model & $w_{log}$CDM model & $w_{log}$CDM model \\
& \textcolor{magenta}{$\Lambda$CDM} & \textcolor{magenta}{$\Lambda$CDM} & \textcolor{magenta}{$\Lambda$CDM} \\
\hline
$H_0\,[{\rm km}/{\rm s}/{\rm Mpc}]$ & $65.5^{+2.5}_{-2.9}$ & $67.23\pm0.84$ & $71.02\pm0.66$ \\
& \textcolor{magenta}{$68.75\pm0.69$} & \textcolor{magenta}{$68.18\pm0.67$} & \textcolor{magenta}{$70.59\pm0.56$} \\
\hline
$\Omega_{\rm m}$ & $0.311\pm0.024$ & $0.2944\pm0.0087$ & $0.2863\pm0.0080$ \\
& \textcolor{magenta}{$0.287\pm0.014$} & \textcolor{magenta}{$0.304\pm0.011$} & \textcolor{magenta}{$0.311\pm0.012$} \\
\hline
$M_B$ & $0$ & $-19.430\pm0.024$ & $-19.353\pm0.017$ \\
& \textcolor{magenta}{$0$} & \textcolor{magenta}{$-19.341\pm0.022$} & \textcolor{magenta}{$-19.375\pm0.022$} \\
\hline
$w_0$ & $-0.68\pm0.21$ & $-0.845\pm0.058$ & $-0.875\pm0.066$ \\
& \textcolor{magenta}{$0$} & \textcolor{magenta}{$0$} & \textcolor{magenta}{$0$} \\
\hline
$w_a$ & $-0.82^{+0.66}_{-0.54}$ & $-0.36^{+0.33}_{-0.26}$ & $-0.69^{+0.37}_{-0.32}$ \\
& \textcolor{magenta}{$0$} & \textcolor{magenta}{$0$} & \textcolor{magenta}{$0$} \\
\hline
\end{tabular}}
\end{table*}

\begin{figure*}[hbt!]
    
  \includegraphics[width=1.0\linewidth]{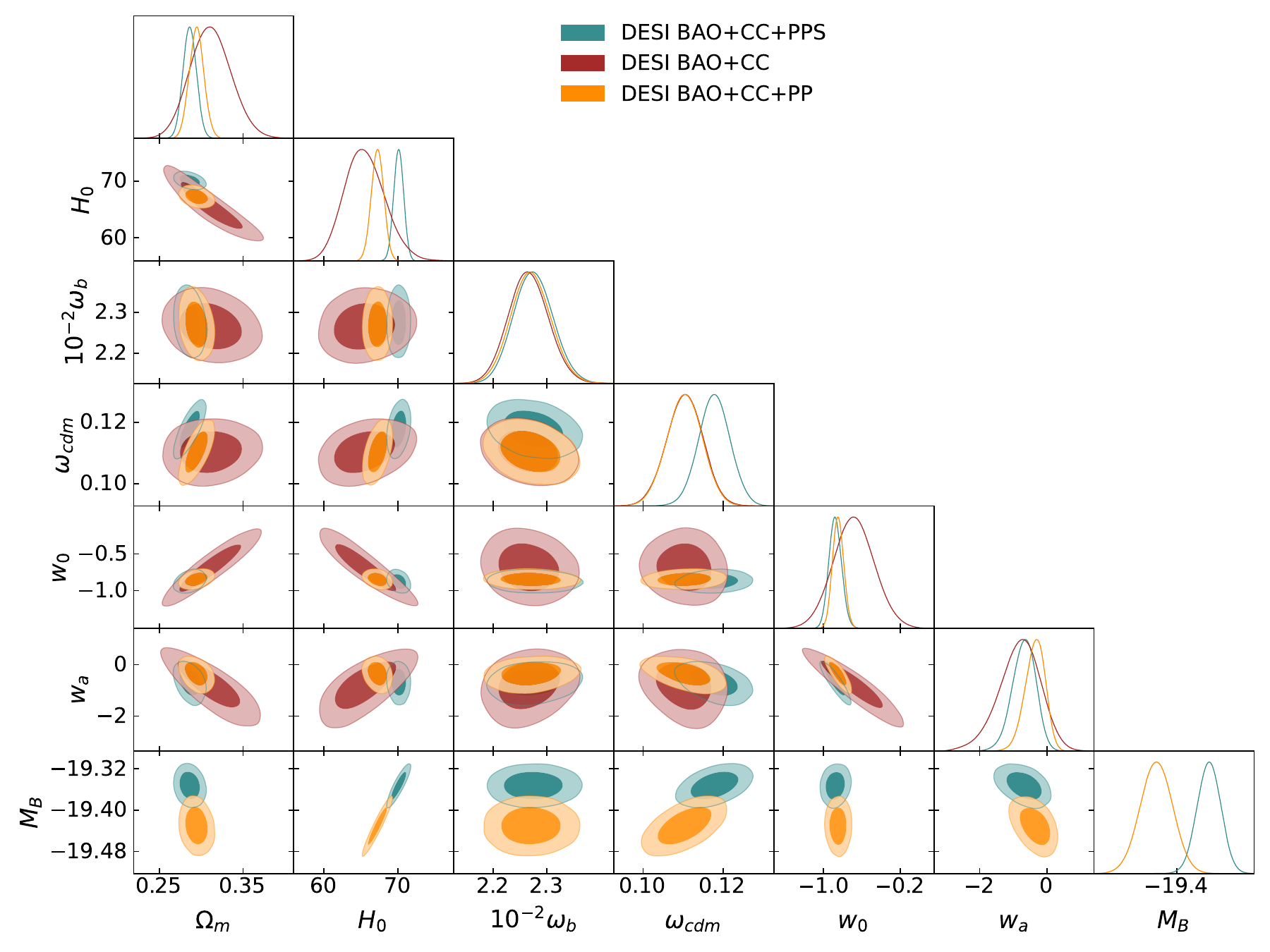}

    \caption{One-D posterior distributions and Two-D marginalized confidence regions ($68\%$ CL and $95\%$ CL) for  $\Omega_{\rm m}$, $M_B$, $w_0$, $w_a$ $10^{-2}\omega_{b}$ and $H_0$ obtained from the DESI BAO+BBN+CC ,DESI BAO+BBN+CC+PP and DESI BAO+BBN+CC+PPS for the $w_{log}$CDM model. The parameter $H_0$ is in units of $km/s/Mpc$.}
    \label{f3}
\end{figure*}

Figure \ref{f3} presents the full parameter space exploration in the $w_{log}$CDM model, showing the correlations among $w_0$, $w_a$, $H_0$, $\Omega_m$, and $M_B$. The corner plot reveals several notable features: (i) a strong negative correlation between $w_0$ and $M_B$, which is expected because these parameters jointly determine the distance-luminosity relation for supernovae; (ii) weak correlations between the dark energy parameters ($w_0$, $w_a$) and the matter density parameters, suggesting that the geometry and expansion history provide complementary constraints; (iii) the Hubble constant shows modest correlations with both $w_0$ and $w_a$. The one-dimensional posteriors indicate preference for slight phantom dark energy ($w_0<-1$) with moderate evidence for non-zero evolution ($w_a\neq0$). Notably, an upward shift in  inferred value of 
$H_0$ does not originate solely from dark-energy dynamics. Part of this shift reflects the well-known degeneracies among $H_0$, the dark-energy equation-of-state parameters and the supernova absolute magnitude $M_B$, particularly when the PPS dataset is included. The strong correlation between $M_B$ and $H_0$, visible in the joint posterior distributions, indicates that supernova calibration plays a significant role in driving the preferred higher values of $H_0$, in addition to the freedom introduced by the dynamic dark-energy parameters.

\begin{figure}[hbt!]
    
  \includegraphics[width=0.5\linewidth]{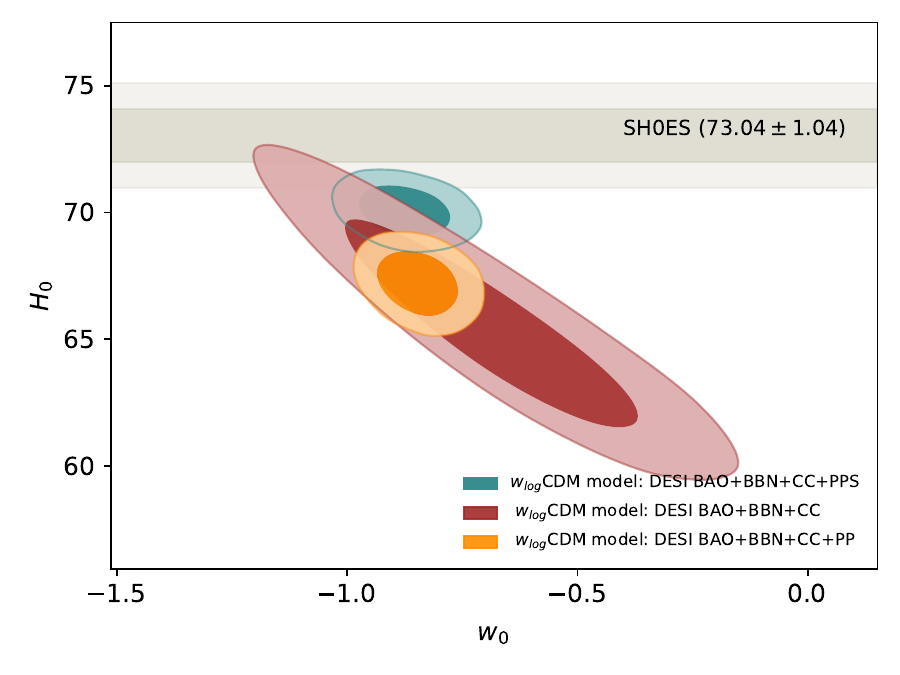}

    \caption {Two–dimensional marginalized confidence contours in the $w_{0}$–$H_{0}$ plane for the $w_{log}$CDM model using the DESI BAO+BBN+CC, DESI BAO+BBN+CC+PP, and DESI BAO+BBN+CC+PPS dataset combinations. The shaded horizontal band shows the SH0ES local determination of the Hubble constant, $H_{0}=73.04 \pm 1.04~\mathrm{km\,s^{-1}\,Mpc^{-1}}$. }

    \label{f4}
\end{figure}

Figure \ref{f4} shows the $w_0 - H_0$ plane, which  presents a clear positive correlation that is expected theoretically expectations and previous observational studies. Models with more negative $w_0$ (phantom dark energy) tend to
favor lower values of $H_0$, whenever those approaching $w_0=-1$ (quintessence) allow for higher $H_0$ values. approaching  local measurements. This degeneracy has been extensively discussed in recent literature \cite{ref82,ref83} as a potential avenue for alleviating  Hubble tension. However, our results indicate that while dynamical dark energy models can shift $H_0$ upward relative to $\Lambda$CDM, the strongest constraints from DESI BAO+BBN+CC+PPS yield $H_0$ values clustering at $\sim$70 km s$^{-1}$ Mpc$^{-1}$, with posteriors predominantly below 70 km s$^{-1}$ Mpc$^{-1}$, suggesting that if the SH0ES measurement is accurate, additional new physics beyond time-varying dark energy may be required.

\begin{figure}[hbt!]
    
  \includegraphics[width=0.5\linewidth]{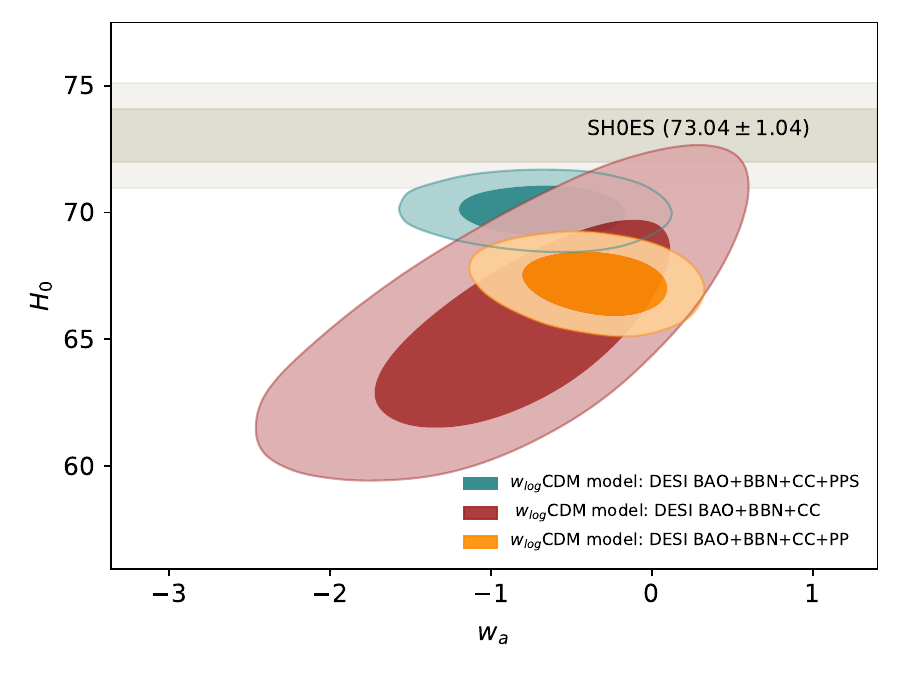}

    \caption{{Two–dimensional marginalized confidence contours in the $w_{a}$–$H_{0}$ plane for the $w_{log}$CDM model using the DESI BAO+BBN+CC, DESI BAO+BBN+CC+PP, and DESI BAO+BBN+CC+PPS dataset combinations. The shaded horizontal band shows the SH0ES local determination of the Hubble constant, $H_{0}=73.04 \pm 1.04~\mathrm{km\,s^{-1}\,Mpc^{-1}}$. }}
    \label{f5}
\end{figure}

 The correlation between $w_a$ and $H_0$ is shown in Figure \ref{f5}. There is a clear positive correlation between these parameters: as $w_a$ increases, indicating a more substantial evolution in the dark energy equation of state, the value of the Hubble constant also tends to increase. As can be seen, this relation is important in the context of the ongoing Hubble tension \cite{ref82,ref83}, where local measurements from the SH0ES collaboration \cite{ref81} provide $H_0 = 73.04 \pm 1.04$ km s$^{-1}$ Mpc$^{-1}$ , which is significantly higher compared to estimates from early universe observations. Our analysis shows that time-varying dark energy models can accommodate higher values of $H_0$ compared to the standard $\Lambda$CDM model. In any case, central values remain closer to the CMB-preferred range of $67-68$ km s$^{-1}$ Mpc$^{-1}$. The DESI BAO+BBN+CC+PPS combination yielded the most restrictive bounds, with $H_0$ values mostly falling in the $65-70$ km s$^{-1}$ Mpc$^{-1}$ range.\\

These correlations illustrate that late-time dynamical dark-energy models can shift the inferred value of $H_0$ upward relative to $\Lambda$CDM. Nevertheless, the resulting values remain largely below the SH0ES determination, reinforcing the conclusion that such models can only partially alleviate the Hubble tension. Our findings are consistent with several previous studies showing that phenomenological dark-energy parameterizations can partially alleviate the Hubble tension without fully resolving it. A similar behavior has been reported in dynamical dark-energy analyses employing extended equation-of-state parameterizations and late-time modifications of the expansion history (see Ref. \cite{add1,add2,add3,add4}). These works support the view that late-time physics alone may be insufficient to completely reconcile the local and early-universe measurements of the Hubble constant.
\begin{figure*}[hbt!]
    
  \includegraphics[width=0.55\linewidth]{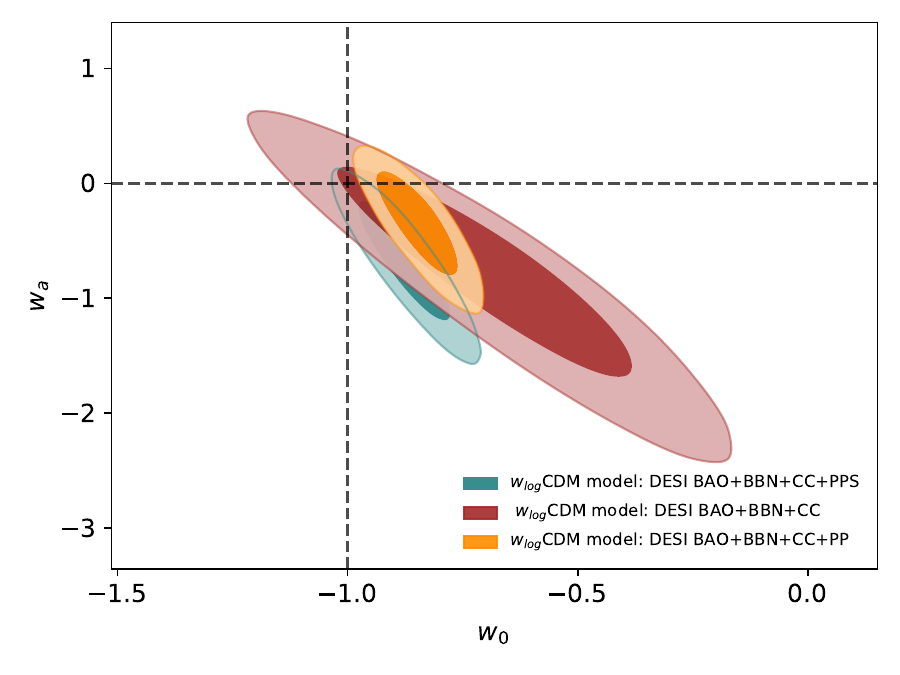}

    \caption{Two-dimensional marginalized confidence regions at 68\% and 95\% C.L of $w_0$ and $w_a$ for the $w_{log}$CDM}
    \label{f6}
\end{figure*}

Figure \ref{f6}  shows the two-dimensional marginalized contours in the $w_0-w_a$ plane at $68\%$ and $95\%$ confidence levels for the three dataset combinations. The constraints show a characteristic negative correlation between $w_0$ and  $w_a$, consistent with previous analyses from the Planck \cite{ref84} and Dark Energy Survey collaborations \cite{ref85}. This is a degeneracy because different combinations of these two parameters yield almost the same expansion history within the observed redshift range. When Pantheon Plus supernova data are included, namely DESI BAO+BBN+CC+PPS, the constraints become tighter, significantly decreasing the allowed parameter space when compared to the configuration excluding supernovas. Our results show that $w_0$ ranges from approximately $-1.5$ to $-0.5$, while $w_ a$ ranges from  $-2$ to $+1$, allowing both quintessence $(w>-1)$ and phantom $(w<-1)$ dark energy models within the uncertainty range.\\

\begin{figure*}[hbt!]
    
  \includegraphics[width=0.5\linewidth]{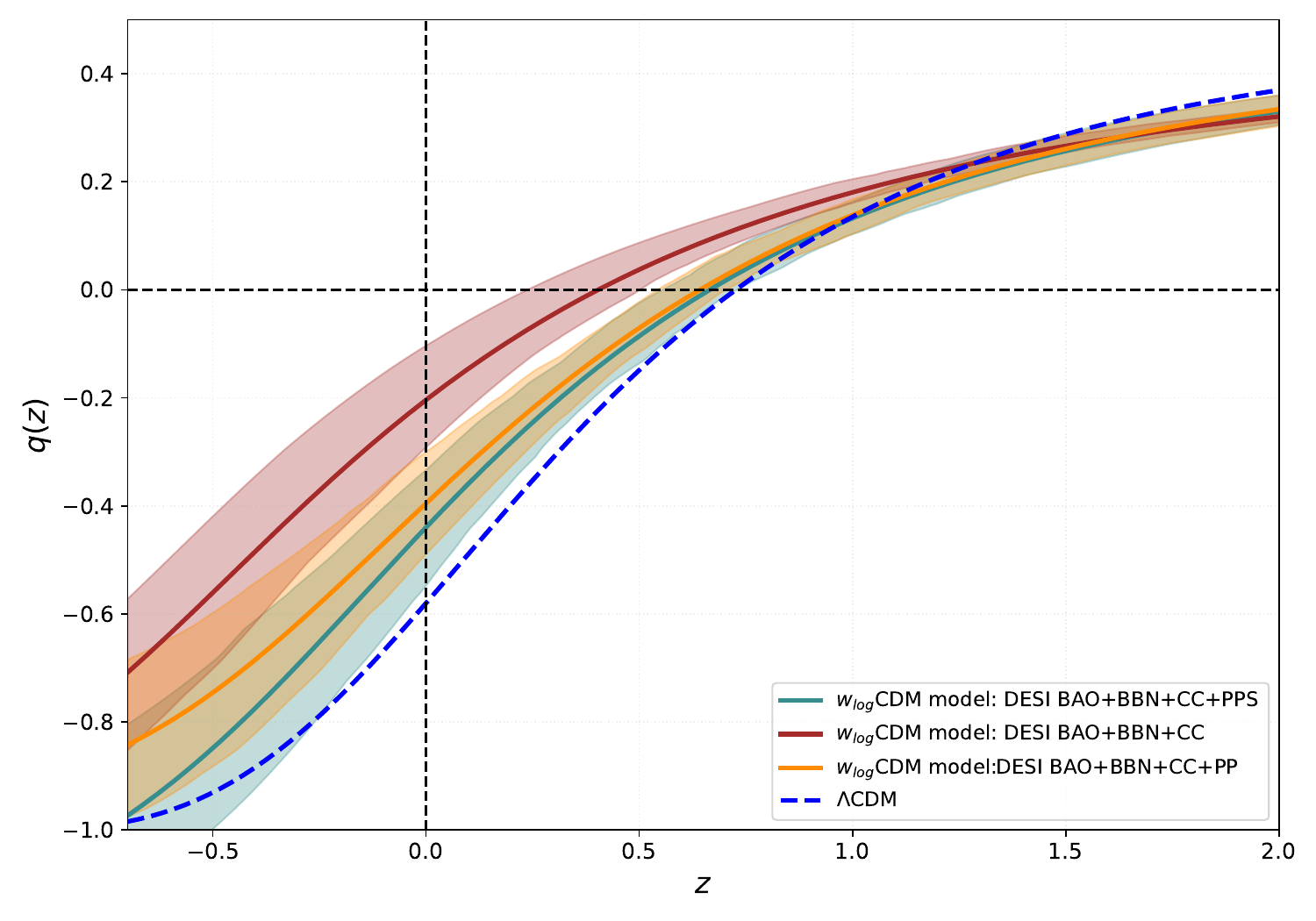}

    \caption{Evolution of the Deceleration parameter $q$ as a function of redshift $z$ for the $w_{log}CDM$ model}
    \label{f7}
\end{figure*}

\textbf{Deceleration Parameter Evolution:-}
The deceleration parameter is a model-independent measure of cosmic acceleration given by $q(z) = -1 - \dot H/H^2 = -1 + (1+z)(1/H)(dH/dz)$; accelerated expansion corresponds to the region where $q<0$, and deceleration is represented when $q>0$. Figure \ref{f7} presents the reconstructed $q(z)$ evolution in the combined data sets with the $w_{log}$CDM model, in which the $\Lambda$CDM prediction is the reference curve. At the present epoch ($z=0$), all three combinations yielded negative deceleration parameters, confirming the accelerated expansion of the universe discovered by the Supernova Cosmology Project and the High-$z$ Supernova Search Team in 1998  \cite{ref1,ref2}. The $\Lambda$CDM model predicts a smooth transition from deceleration to acceleration at $z \sim 0.6 - 0.7$, marked by  $q=0$ \cite{ref86}. Our reconstructions show reasonable agreement with this prediction, though with varying degrees of precision depending on the dataset combination. In particular, the DESI BAO+BBN+CC result had the largest uncertainties, with error bands at low and intermediate redshifts extending from strongly negative values of $q \sim -1$ to mildly positive values. The combination of DESI BAO+BBN+CC+PPS provides the most precise determination, with narrow error bands, which strongly supports the standard model at $z < 1.5$. At higher redshifts ($z>1$), all three reconstructions converge toward positive $q$ values, consistent with the matter-dominated epoch where gravitational attraction caused cosmic deceleration before dark energy became dominant \cite{ref87}. Importantly, the redshift of transition for which $q=0$ obtained from our analysis is compatible, within uncertainties, with the expected value from $\Lambda$CDM at $z \sim 0.5 - 0.8$ \cite{ref88}. The reconstruction of $q(z)$ showed a relatively smooth evolution across all datasets without sharp features or discontinuities, signifying the robustness of the reconstructions. These reconstructions show that dynamical dark energy models succeed in capturing the transition from early-time deceleration to late-time acceleration, although precise determination of the transition redshift and the behavior at $z > 1$
requires the integrated power of several complementary datasets \cite{ref89}.

\begin{figure*}[hbt!]
    
  \includegraphics[width=0.5\linewidth]{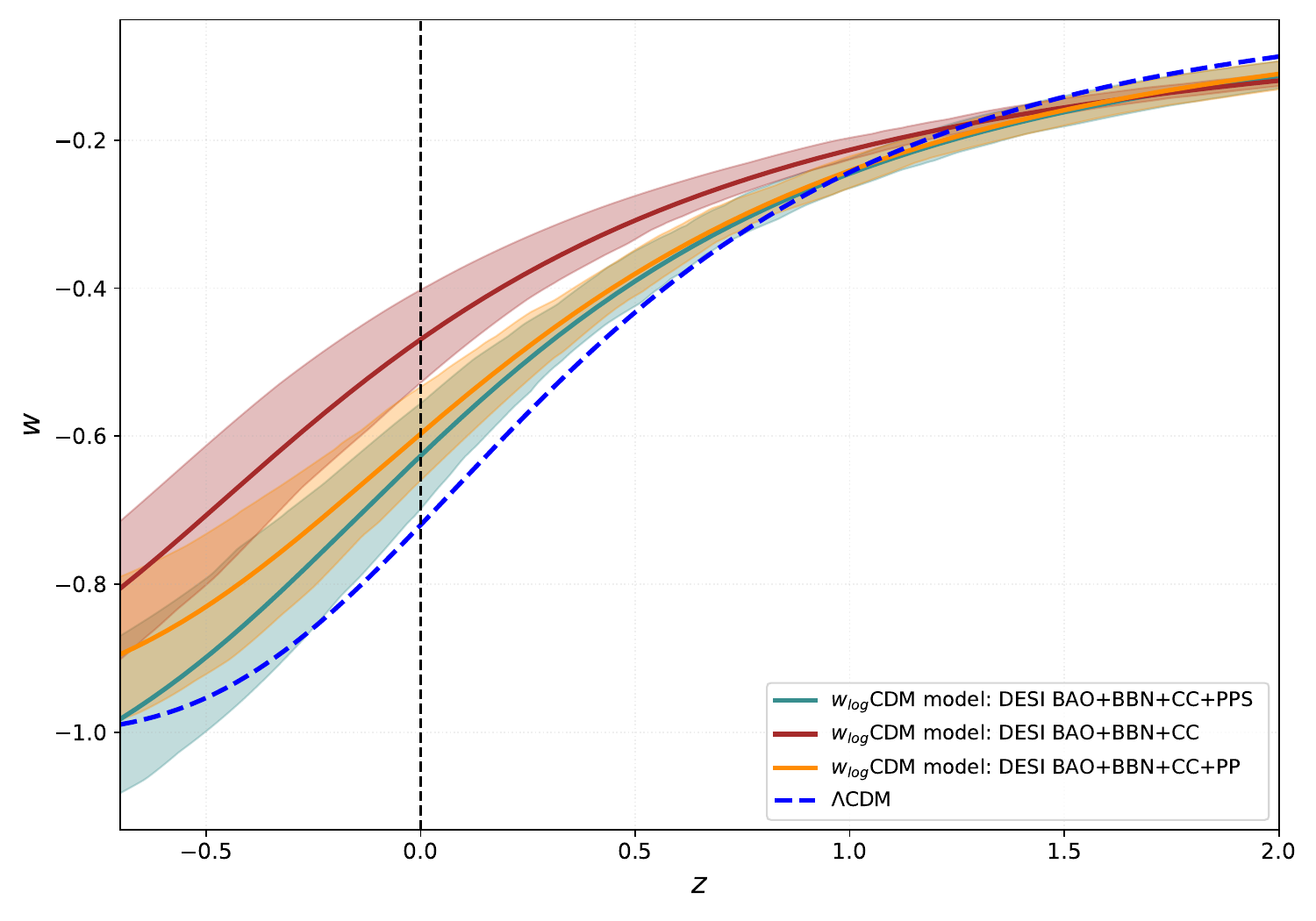}

    \caption{Evolution of the dark energy equation of state parameter $w$ as a function of redshift $z$ for the $w_{log}CDM$ model.}
    \label{f8}
\end{figure*}

\textbf{Evolution of the Equation of State parameter with Redshift:-}
Figure \ref{f8} Reconstructed evolution of the dark energy equation of state parameter $w_{d}(z)$ as a function of redshift by the $w_{log}$CDM model for the three dataset combinations, in comparison to the standard $\Lambda$CDM prediction of $w=-1$. The equation of state parameter is a phenomenological description of the behavior of dark energy. In this figure, the reconstruction shows how the effective equation of state changes across cosmic time, from the present epoch ($z=0$) to higher redshifts ($z \sim 2$). Qualitatively, all three dataset combinations are rather similar, with $w(z)$ starting near or below $-1$ in the present and varying at higher redshifts. The DESI BAO+BBN+CC combination had the largest uncertainties, as reflected by the broader shaded region, owing to the small constraining power without supernova data. Adding Pantheon Plus(DESI BAO+BBN+CC+PP) significantly reduces these uncertainties, especially for intermediate redshifts around $z \sim 0.5 - 1.0$, where the BAO measurements provide strong geometrical constraints \cite{Demianski2020}. DESI BAO+BBN+CC+PPS is the most constraining combination, because of the high-redshift leverage offered by Pantheon Plus supernovae extending up to $z \sim 2$ \cite{Demianski2013}. Notably, all three reconstructions are consistent with the cosmological constant value of $w=-1$ within their respective uncertainty bands. However, the
central trajectories suggest mild deviations toward phantom behavior ($w<-1$) at low redshifts and possible evolution at $z>1$. The $w_{log}$CDM model allows for more flexible behavior compared to the CPL parameterization, capturing potential non-monotonic evolution in the equation of state that may be present in realistic scalar field models \cite{Crittenden2012}. The overlap of different combinations of datasets at $z < 0.5$ indicates a good degree of agreement between independent probes of the well-constrained local universe and the declining overlaps at higher redshifts illustrate at which higher redshifts the datasets provide complementary information. Thus, the results indicate that the available datasets cannot rule out mildly dynamical dark energy models, although the data are generally consistent with the cosmological constant over the redshift coverage. The DESI data provided clear evidence of the evolution of dark energy of moderate significance.

\section{Model Comparison}
The current subsection concludes the observational analysis by comparing the fits of the different models based on accepted information criteria. The two primary examples of such criteria are the Akaike Information Criterion (AIC) \cite{refa} and the Bayesian or Schwarz Information Criterion (BIC) \cite{refb}. These are  described as follows:
\begin{eqnarray}
&& AIC  = -2 \ln \mathcal{L}+ 2 d = \chi^2_{min} + 2 d,\label{aic}
\end{eqnarray}
and
\begin{eqnarray}
&& BIC  = -2 \ln \mathcal{L}+  d \ln N = \chi^2_{min} +  d \ln N,\label{bic}
\end{eqnarray}

Here, $\mathcal{L} = \exp \left(-\chi_{\text{min}}^2/2\right)$ is the maximum likelihood, where $d$ is the number of parameters in the model and $N$ is the total number of data points considered in the analysis. To compare the different models, it is necessary to use a reference model, which is the standard $\Lambda$CDM cosmology. The performance of a given model $M$, its performance can be evaluated via the difference $\Delta X = X_M - X_{\Lambda\text{CDM}}$, where $X$ can take either the AIC or BIC. The value of $\Delta X$, it provides the following support for a model:
i) strongest evidence if $\Delta X \leq 2$ (a good fit),
(ii) moderate support if $4 \leq \Delta X \leq 7$, and
(iii) essentially no support if $\Delta X \geq 10$. \\
The computed values of AIC, BIC, and $\chi^2$ for our models are presented in Section \ref{t2}, where we compare them with the standard $\Lambda$CDM cosmology using three  observational datasets: DESI BAO+BBN+CC, DESI BAO+BBN+CC+PP, and the combined DESI BAO+BBN+CC+PPS. We also report the relative differences $\Delta$AIC and $\Delta$BIC with respect to the base $\Lambda$CDM model to ensure a fair comparison. We find that the AIC and BIC values of the $w_{\log}$CDM model are systematically but only marginally higher than those of $\Lambda$CDM \cite{refc,refd,refe} for all dataset combinations. This indicates that, although the extended model provides a phenomenologically viable and statistically competitive description of the data, the improvement in the goodness of fit does not compensate for the presence of additional free parameters according to standard information-criterion analyses. Consequently, the current data do not statistically prefer the $w_{\log}$CDM scenario over $\Lambda$CDM, which remains the favored model.
.\\

\begin{table*}[hbt!]

\label{t2}
 \resizebox{\textwidth}{!}{
\begin{tabular} { |c | c| c| c|l|l|  }   
 \hline

Data

 & DESI BAO+BBN+CC &   DESI BAO+BBN+CC+PP & DESI BAO+BBN+CC+PPS \\

 \hline
 Model      & $w_{log}$CDM model  &  $w_{log}$CDM model    & $w_{log}$CDM model             \\
      & \textcolor{magenta}{$\Lambda$CDM}  &  \textcolor{magenta}{$\Lambda$CDM}     & \textcolor{magenta}{$\Lambda$CDM}       \\
\hline

$AIC$     &        $43.92     $ &$1453.92   $ & $1357.30  $  \\

    &  \textcolor{magenta}{$40.16$}
& \textcolor{magenta}{$1450.76$}    &  \textcolor{magenta}{$1354.12$}
\\

\hline
$\Delta AIC$&             $    3.76         $ &$   3.16      $ &  $3.18            $ 
\\
 
    &  \textcolor{magenta}{$0$}
& \textcolor{magenta}{$0$}    &  \textcolor{magenta}{$0$}
\\

\hline

$BIC$          & $  53.91       $ &$   1479.75      $ &  $   1383.16         $ 
\\
 
    &  \textcolor{magenta}{$45.15 $}
& \textcolor{magenta}{$1469.94$}    &  \textcolor{magenta}{$ 1377.04$}
\\
\hline

$\Delta BIC$&     $     8.76      $ &$  9.81     $ &  $ 6.12  $ 
\\

   &  \textcolor{magenta}{$0$}
& \textcolor{magenta}{$0$}    &  \textcolor{magenta}{$ 0 $}
\\
\hline

$\chi^{2}_{min}$&     $     39.92      $ &$  1449.92      $ &  $ 1353.30  $ 
\\

   &  \textcolor{magenta}{$38.16$}
& \textcolor{magenta}{$1448.76$}    &  \textcolor{magenta}{$ 1352.12 $}
\\
\hline

\end{tabular}
}

\caption{
A summary of the $AIC$ and $BIC$ values, along with their deviations from the standard $\Lambda$CDM cosmological model and $w_{log}$CDM model. The analysis includes different data combinations: DESI BAO+BBN+CC, DESI BAO+BBN+CC+PP, and DESI BAO+BBN+CC+PPS.}
\vspace{0.5cm}
\label{t2}
\end{table*}

\section{Conclusion}

In this study, we investigated the logarithmic dark energy model, denoted as the $w_{\log}$CDM scenario, characterized by the equation of state $$w_{d}(z) = w_{0} + w_{a}\left( \frac{\ln(2+z)}{1+z} - \ln 2 \right),$$ which allows for a mild, smooth evolution of the dark energy sector as a function of the redshift. By combining the latest cosmological datasets—including DESI BAO, BBN, Cosmic Chronometers, and the Pantheon Plus supernova sample (with and without SH0ES calibration), we  placed stringent observational constraints on the parameters of this model and examined their implications for the expansion history of the universe. Using the full dataset combination DESI BAO+BBN+CC+PPS, we obtained the constraints $H_{0} = 71.02 \pm 0.66~\mathrm{km\,s^{-1}\,Mpc^{-1}}$, $\Omega_{m}=0.2863 \pm 0.0080$, $w_{0} = -0.875 \pm 0.066$, and  $w_{a} = -0.69^{+0.37}_{-0.32}$ 
(68\% and 95\% CL). These results indicate a mild preference for phantom-like dark energy ($w<-1$), which remained statistically compatible with a cosmological constant within $2\sigma$. The negative evolution parameter $w_a$ further hints at a possible redshift dependence of the dark energy equation of state, although the statistical significance remains limited.\\

One of the notable outcomes of our analysis was the Hubble constant. The $w_{\log}$CDM model yields an $H_0$ value that is significantly higher than the prediction of the standard $\Lambda$CDM model inferred from the early universe datasets. When the PPS (Cepheid-calibrated) sample is included, the inferred value of $H_{0}$ approaches the local SH0ES determination, thereby partially reducing the long-standing Hubble tension. Although a residual $\sim 2\sigma$ discrepancy persists, our results demonstrate that dynamical dark energy models with mild evolution, such as the $w_{\log}$CDM scenario, are capable of shifting the inferred expansion rate toward the locally measured values more effectively than constant-$w$ or $\Lambda$CDM models. The reconstructed deceleration parameter, $q(z)$, exhibits a transition from decelerated to accelerated expansion at redshift $z \sim 0.6$--$0.7$, which is in excellent agreement with previous observational studies. Similarly, the reconstructed dark energy equation of state remains close to $w=-1$ across the redshift range probed by the current data, with deviations consistent with a mild dynamical evolution. These findings highlight the capability of the logarithmic parameterization to capture subtle departures from the cosmological constant without introducing extreme or pathological behavior at low or high redshift. The parameter-space analysis reveals several physically meaningful correlations. In particular, $H_0$ shows a positive degeneracy with $w_0$ and $w_a$, consistent with theoretical expectations that phantom-like dark energy tends to push the inferred Hubble constant downward, whereas models closer to $w=-1$ accommodate higher values of $H_0$. The correlations between $M_B$, $w_0$, and $H_0$ further demonstrate the importance of distance-ladder calibration in shaping constraints on dynamical dark energy. Information-criterion tests indicate that the $w_{\log}$CDM model performs comparably to $\Lambda$CDM, showing no statistically significant preference for added complexity, but also no penalty strong enough to disfavor the model. Finally, we stress that the $_{log}$CDM scenario explored in this study should be regarded as an effective phenomenological extension of the standard cosmological model. Although it does not provide a fundamental explanation for the physical origin of dark energy, it offers a flexible and observationally testable framework to investigate possible late-time deviations from $\Lambda$CDM. \\

Overall, our results suggest that the $w_{\log}$CDM model provides a viable and competitive extension of the standard cosmological framework, capable of capturing possible dynamics in the dark energy sector while remaining consistent with current observational data. While the $w_{log}$CDM model leads to a statistically significant upward shift in the inferred value of the Hubble constant, the remaining $\sim 2\sigma$ discrepancy with the SH0ES measurement indicates that the Hubble tension is alleviated but not resolved. Consistent with previous studies on late-time dynamical dark-energy models, it meaningfully reduces its significance and offers a well-motivated phenomenological avenue for further exploration. Future high-precision surveys—such as DESI Year 3, Roman Space Telescope, and Euclid—will be invaluable in tightening constraints on the dynamical evolution of dark energy and determining whether mild departures from $\Lambda$CDM, such as those predicted by the logarithmic parameterization, are favored by next-generation observations. Finally, we emphasize that, from the perspective of information-criterion analyses based on AIC and BIC, the $w_{\log}$CDM model should be regarded as a statistically competitive phenomenological extension of the standard cosmological scenario. However, the current observational data do not indicate a statistical preference for this extended model over $\Lambda$CDM, which remains the favored description of the late-time Universe.

\section*{Declaration of competing interest}
The authors declare that they have no known competing financial
interests or personal relationships that could have  influenced
the work reported in this study.

\section*{Data availability statement}
We employed publicly available Pantheon Plus, Pantheon Plus SH0ES, Cosmic Chronometer (CC), BBN  and DESI BAO data presented in this study. The CC data were compiled from publicly available cosmic chronometer measurements in the literature, with a representative compilation accessible at: https://github.com/AhmadMehrabi Cosmic\_chronometer\_data. The Pantheon Plus, Pantheon Plus SH0ES and BBN data compilation (distance moduli and covariance matrices),  which is publicly available on GitHub: https://github.com/brinckmann/montepython\_public/tree/3.6/montepython/likelihoods and the DESI BAO data which are publicly available on Github: https://github.com/LauraHerold/MontePython\_desilike/blob/main/likelihood/bao\_desi\_all/ . No additional data we are used in this study.

\section*{acknowledgments}
The authors (A. Dixit and A. Pradhan) thank the IUCAA, Pune, India for providing the facility under visiting associateship program. The author (S. Verma) was supported by a Senior Research Fellowship (UGC Ref No. 192180404148) from the University Grants Commission, Govt. of India.

\end{document}